\documentstyle[prl,aps,psfig,multicol]{revtex}
\begin{document}
\title{Fate of the extended states in a vanishing magnetic field:\\ the
role of spins in strongly-interacting 2D electron systems}
\author{M.~R. Sakr, Maryam Rahimi, and S.~V. Kravchenko}
\address{Physics Department, Northeastern University, Boston,
MA 02115, U.S.A.}
\date{\today}
\maketitle
\begin{abstract}
In non-interacting or weakly-interacting 2D electron systems, the energy of the extended states increases as the perpendicular magnetic field approaches zero: the extended states ``float up" in energy, giving rise to an insulator.  However, in those 2D systems where metallic conductivity has been recently observed in zero magnetic field, the energy of the extended states remains constant or even decreases as $B_\perp\rightarrow0$, thus allowing conduction in the limit of zero temperature.  Here we show that aligning the electrons' spins causes the extended states to once more ``float up" in energy in the vanishing perpendicular magnetic field, as they do for non- or weakly-interacting electrons.  The difference between extended states that float up (an insulator) or remain finite (a metal) is thus tied to the existence of the spins.
\end{abstract}
\begin{multicols}{2}

When a strong perpendicular magnetic field is applied to a two-dimensional electron system, extended (delocalized) states exist at semi-integer filling factors of Landau levels, $\nu\equiv n_s\, (ch/eB_\perp)=i/2$.  Here $n_s$ is the electron density, $c$ is the speed of light, $h$ is the Plank's constant, $e$ is the electron charge, $B_\perp$ is the component of the magnetic field perpendicular to the 2D plane, and $i=$~1, 2, 3....  The extended states separate different quantum Hall states with vanishing longitudinal conductivity, $\sigma_{xx}=0$, and quantized Hall conductivity, $\sigma_{xy}=i\, e^2/h$.  The lowest extended state separates the quantum Hall state with $\sigma_{xx}=0$ and $\sigma_{xy}=1\, e^2/h$ at higher energies and the insulating state with $\sigma_{xx}=\sigma_{xy}=0$ at lower energies.  In the absence of a magnetic field, however, noninteracting electrons are always localized in 2D \cite{abrahams79}.  The fate of the finite-field extended states as the magnetic field approaches zero has been a subject of debate for almost 20 years.  The prevailing view holds that the energy of these extended states increases to infinity as the field approaches zero, giving rise to an insulator \cite{khmelnitskii84,laughlin84,kivelson92,glozman95}.  However, the situation is different in {\em strongly interacting} electron systems where metallic conductivity has been recently observed in zero magnetic field \cite{abrahams00}.  In such systems, the energy of the extended states remains finite, thus allowing conduction in the limit of zero temperature \cite{pudalov93,shashkin93,dultz98}.  Here we show that aligning the electrons' spins causes the extended states to once more increase (``float up'') in energy in a vanishing perpendicular magnetic field, as they do for non-interacting electrons.  The difference between extended states that increase in energy (an insulator) or remain finite (a metal) is thus unambiguously tied to the existence of the spins.

The 2D electron gas was studied in low-disordered (maximum mobility of about 2.5~m$^2$/Vs) silicon metal-oxide-semiconductor field-effect transistors (MOSFETs).  Different methods have been used in the past to map out the positions of the extended states.  Here we use two criteria previously employed by Glozman~{\it et al.} \cite{glozman95} and Shashkin~{\it et al.} \cite{shashkin93} which give similar results.  The position of the lowest extended state was determined by requiring that the nonlinearity of the current-voltage ($I$-$V$) characteristics vanish \cite{shashkin93}.  The $I$-$V$ curves are strongly nonlinear in the insulating state where the resistance exponentially diverges as temperature is decreased; an example is shown in Fig.\ref{IV}.  A typical $I$-$V$ curve, measured at low temperature, resembles a step-like function with a voltage that rises abruptly by twice a threshold value, $2V_{th}$, as the current passes through zero.  The square root of $2V_{th}$ is plotted as a function of $n_s$ for each $B_\perp$, as shown in the inset.  The extrapolation of the $(2V_{th})^{1/2}(n_s)$ dependence to zero threshold value yields the ``critical'' electron density at which both the nonlinearity and the exponential divergence of the resistance as $T\rightarrow0$ vanish, and the states are extended.  For more about the procedure and its justification, see Refs.\cite{shashkin93,shashkin00}.  To map out the positions of higher extended states, {\it i.e.}, those between neighboring quantum Hall effect states, we traced maxima of the diagonal conductivity, as was first done by Glozman {\it et al.}\cite{glozman95}.  To polarize the electrons' spins, we apply a constant magnetic field of 6~Tesla.  At the low temperatures and low electron densities used in our experiments, this magnetic field is sufficiently strong to cause full spin polarization.  This was checked by measuring the dependence of the resistance on magnetic field $B_{||}$ parallel to the 2D plane.  At all electron densities used in our experiments, this dependence saturates at $B_{||}\lesssim4$~Tesla; the saturation signals the onset of the full spin polarization \cite{okamoto99,vitkalov00}.  To map out the positions of the extended states on a $(B_\perp,n_s)$ plane for spin-unpolarized electrons \cite{unpolarized}, we applied a magnetic field perpendicular to the 2D plane.  For spin-polarized electrons, we tilted the sample in a constant magnetic field of 6~Tesla with the help of a rotator thus producing a perpendicular component of the field.

Our experimental results are shown as data points in Fig.\ref{diagram}.  The positions of the lowest extended state are denoted by blue circles for spin-unpolarized electrons and by red circles for fully spin-polarized electrons.  Within the accuracy of our experiment, their positions coincide for $B_\perp\gtrsim1.5$~Tesla.  Since full spin polarization is reached 
\vbox{
\vspace{3.0mm}
\hbox{
\psfig{file=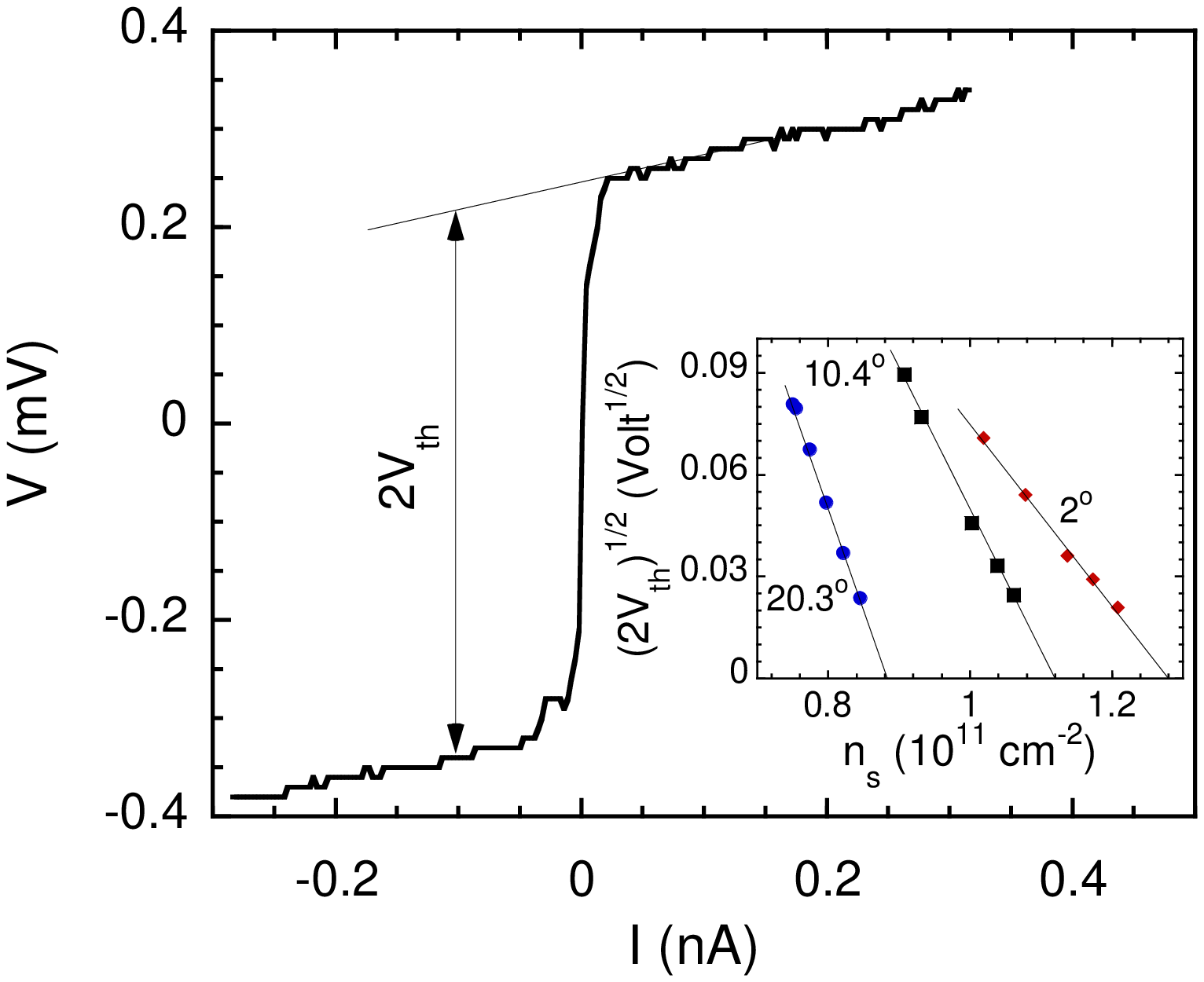,width=3.0in,bbllx=.5in,bblly=1.25in,bburx=7.25in,bbury=9.5in,angle=0}
}
\vspace{-1.4in}
\hbox{
\hspace{-0.15in}
\refstepcounter{figure}
\parbox[b]{3.4in}{\baselineskip=12pt \egtrm FIG.~\thefigure.
Typical current-voltage characteristic of the insulating spin-polarized electron system. $T=35$~mK.  The inset shows the square root of the threshold voltage as a function of the electron density for three tilt angles as labeled.\vspace{0.10in}
}
\label{IV}
}
}
only at $B_\perp\gtrsim4$~Tesla, this coincidence demonstrates that the polarization of the spins does not affect the degree of disorder in the system.  The lowest extended state oscillates as a function of field, exhibiting deep minima at integer Landau filling factors $\nu=$~1 and 2, in agreement with previous measurements \cite{shashkin93,diorio90} (red dotted lines in the figure correspond to integer filling factors, as labeled).  As $B_\perp$ is decreased below approximately $1.6$~Tesla ($\nu=2$), the lowest extended state begins to increase sharply in energy \cite{energy}.
 
The main result of this paper is the qualitative difference in the behavior of the lowest extended state as $B_\perp\rightarrow0$ for spin-unpolarized and spin-polarized electrons.  In the unpolarized case, the lowest extended state decreases monotonically in energy after reaching a maximum at $B_\perp\approx1.5$~Tesla.  A decrease of $B_\perp$ thus causes {\em delocalization} of the spin-unpolarized electrons.  In contrast, when the electron spins are fully polarized, the lowest extended state continues to increase in energy as the field is reduced, going through an additional minimum at $\nu=4$.  Decreasing $B_\perp$ toward zero in a strongly interacting spin-polarized electron system thus promotes {\em localization} and restores the ``floating'' behavior expected for non-interacting electrons \cite{khmelnitskii84,laughlin84,kivelson92,glozman95}.

It is interesting to note, however, that in the spin-polarized system, the position of the lowest extended state, as determined by the method used, approaches a finite electron density $n_c^B\approx1.3\cdot10^{11}$~cm$^{-2}$ as $B_\perp\rightarrow0$ rather than increasing indefinitely.  This may imply that a conducting state is possible even for spin-polarized electrons at $B_\perp=0$ for electron densities above $n_c^B$.  However, the temperature dependence of the resistance of the spin-polarized electron system remains ``insulating-like'' ($dR/dT<0$), although not exponential, even at electron densities far above $n_c^B$ \cite{shashkin00}, in 
\vbox{
\vspace{-9mm}
\hbox{
\psfig{file=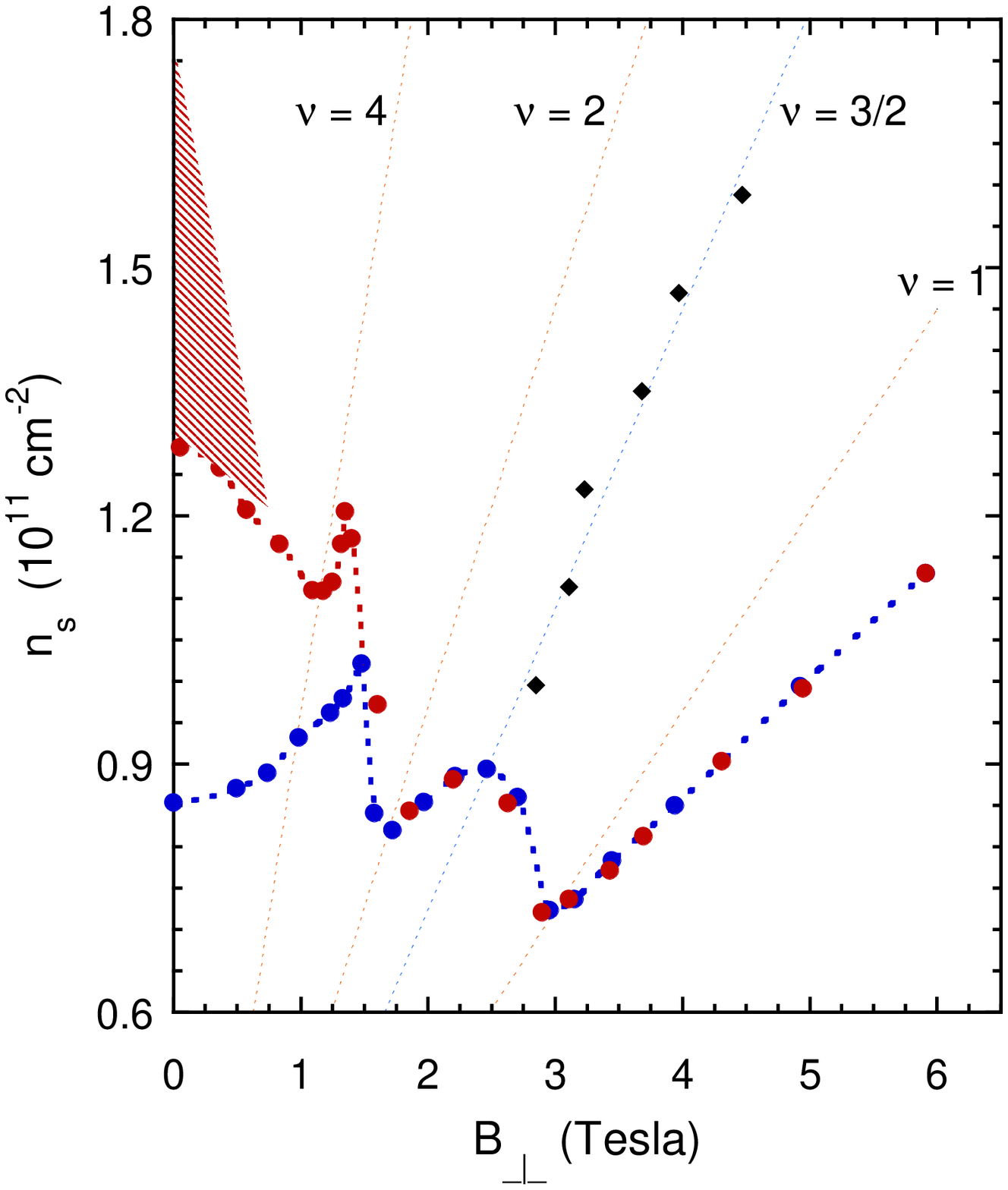,width=3.0in,bbllx=.5in,bblly=1.25in,bburx=7.25in,bbury=9.5in,angle=0}
}
\vspace{0.07in}
\hbox{
\hspace{-0.15in}
\refstepcounter{figure}
\parbox[b]{3.4in}{\baselineskip=12pt \egtrm FIG.~\thefigure.
Positions of the lowest extended states on a $(B_\perp,n_s)$ plane for spin-polarized (red circles) and unpolarized (blue circles) electrons.  The black diamonds show the experimentally determined position of the next extended state between two neighboring quantum Hall states at $\nu=$~1 and 2.  The dotted lines correspond to integer filling factors as labeled (red lines) and to $\nu=3/2$ (blue line).  The shaded area shows the region of parameters where the temperature dependences of the resistance of spin-polarized electron system are ``insulating-like'' ($dR/dT<0$), although non-exponential.\vspace{0.10in}
}
\label{diagram}
}
}
the shaded region in the figure.  This suggests that the spin-polarized electrons are still weakly (non-exponentially) localized in the shaded region and their states are not extended.  (In contrast, the temperature dependence of the resistance of the spin-unpolarized electron system in $B_\perp=0$ is always ``metallic-like'', $dR/dT>0$, at electron densities above the critical one \cite{abrahams00,shashkin00}.)

The positions of several higher extended states were also determined in our experiment.  To simplify the figure, we show only the second lowest extended state (black diamonds) which lies between the quantum Hall states at $\nu=1$ and 2.  In high magnetic fields, it approximately follows the blue dotted line corresponding, as expected, to the center of Landau level ($\nu=3/2$); at a lower field, it appears to merge with the lowest extended state, in agreement with previous experimental \cite{hilke00} and  theoretical \cite{sheng00} results.  We also note that the minimum in the energy of the lowest extended state at $\nu=4$ is not observed in the spin-unpolarized case in accordance with recent magnetoresistance data \cite{kravchenko00}, while it is ``restored'' in the spin-polarized system.

Once their spins are aligned, the electrons can be considered ``spinless'': the spin degree of freedom disappears.  Our results therefore show that it is the existence of the electrons' spins that is responsible for changing the behavior of the extended states in strongly interacting systems: the localization that dominates for noninteracting electrons at $B_\perp\rightarrow0$ is restored once the spin degree of freedom is suppressed.  The question why strong interactions prevent the energy of the extended states from increasing as $B_\perp\rightarrow0$ in spin-unpolarized electron systems remains open.

We are grateful to T. Okamoto and S. Kawaji for providing us with a 
high-mobility Si MOSFET sample.  We would also like to thank M.~P. Sarachik, D.~N. Sheng, and N.~E. Israeloff for helpful discussions and critical remarks on  the manuscript.  S.~V.~K. acknowledges discussions with A.~H. Castro Neto, M.~P.~A. Fisher, and A.~A. Shashkin.  This work was supported by NSF Grants  No.~DMR-9803440 and DMR-9988283 and Sloan Foundation.

\end{multicols}

\begin{thebibliography}{10}
\bibitem{unpolarized} The electrons are referred to as spin-unpolarized if only a perpendicular magnetic field is applied.  However, we keep in mind that any magnetic field results in partial polarization.
\bibitem{energy} We show positions of the extended states on a $(B_\perp,n_s)$ instead of $(B_\perp$,energy) plane.  However, as discussed in Ref.\cite{glozman95}, there is no qualitative difference between these two ways of plotting the data because the energy is approximately proportional to $n_s$ under the conditions of this experiment.
\end{thebibliography}
\end{document}